\documentclass[12pt,cite,epsf,epsfig,psfrag]{article}
\usepackage{epsfig}
\usepackage{amsmath, graphics, setspace}
\usepackage{graphicx}
\usepackage{amssymb}
\usepackage{subfigure}

\font\mybb=msbm10 at 12pt
\def\bb#1{\hbox{\mybb#1}}

\def\T {\bb{T}}

\def\be{\begin{equation}}
\def\ee{\end{equation}}
\def\bea{\begin{eqnarray}}
\def\eea{\end{eqnarray}}

\newcommand{\beqal}{\begin{eqnarray}\label}
\newcommand{\beqa}{\begin{eqnarray}}
\newcommand{\eeqa}{\end{eqnarray}}

\newcommand{\mathsym}[1]{{}}
\newcommand{\unicode}[1]{{}}

\begin{document}

\begin{titlepage}
\begin{center}

\vskip .2in

{\Large \bf Dissipative force on an external quark in heavy quark cloud}
\vskip .5in

{\bf Shankhadeep Chakrabortty}\footnote{e-mail: sankha@iopb.res.in} \\
\vskip .1in
{ Institute of Physics,\\
Bhubaneswar 751~005, India.}
\end{center}

\begin{center} {\bf ABSTRACT}

\end{center}
\begin{quotation}\noindent
\baselineskip 15pt

Within the finite temperature ${\cal N} = 4$ strongly coupled super Yang-Mills,  we
compute the dissipative force on an external quark in the presence of evenly distributed
heavy quark cloud. This is computed holographically by constructing the corresponding gravity
dual. We study the behaviour of this force as a function of the cloud density. Along the way we also 
analyze the stability of the gravity dual for vector and tensor perturbations.
\end{quotation}
\end{titlepage}
\vfill
\eject

Collisions of heavy nuclei are believed to produce QGP, a strongly coupled thermal state of 
matter \cite{Arnold:2007pg,Shuryak:2008eq}. This motivated many researchers to compute
QGP observables using gauge/gravity correspondence. Though the gauge theory in question 
is quite different from QCD, this correspondence gives us a systematic
prescription to compute various quantities associated with  a strongly coupled thermal system 
\cite{Aharony:1999ti}. These include viscosity, entropy production, 
transport coefficients to name a few \cite{Policastro:2001yc, Liu:2006ug, Gubser:2009md}.
In this respect, an interesting quantity is the dissipative force experienced by an
external heavy quark moving in a thermal plasma \cite{Herzog:2006gh, Liu:2006ug,
Gubser:2006bz}. Within the framework of gauge/gravity duality, an external heavy quark is 
modeled by a fundamental string attached to the boundary of an AdS black hole. For ${\cal N}=4$
super Yang-Mills, the end point of the string carry a fundamental $SU(N)$ charge. The string 
extends along the radial direction of the AdS-Schwarzschild metric.
This external quark, with a mass proportional to the length of the string, loses
its energy as the string trails back imparting a drag force.

Our aim in this paper is to carry out a simple calculation of the drag force for
thermal ${\cal N}=4$ super Yang-Mills on $R^3$ in the following scenario. We consider an
{\it uniformly} distributed heavy quark cloud in this hot plasma. We then ask:
how does the drag force on an external quark change with the density of the quark cloud?
Like other drag force calculations, we compute it holographically. First, we construct
the gravity dual of the quark cloud. In the bulk, this represents a black hole
in the presence of a string cloud. These strings, assumed to be non-interacting,
are aligned along the radial direction of the bulk geometry and are distributed
uniformly over $R^3$. We then study the thermodynamics of black hole and verify the thermodynamical stability.
We also analyze it's  gravitational stability 
under vector and tensor perturbation in a gauge invariant way. Subsequently, we introduce a probe string with an end 
on the boundary and calculate the dissipative force on the heavy quark
following the usual approach. At this point, though, the relevance of this work in the light of
recent quarck-gluon-plasma experiment is not immediately obvious. Nonetheless, we would like
to make the following comment. The dynamics of a heavy quark (say charm) passing through
the plasma is usually described by considering it's interaction with the medium and the resulting energy
loss is calculated. In such calculations, any possible effects of other heavy quark due to the 
back-reaction of the plasma are neglected. In the context of ${\cal N} = 4$ SYM, our work can perhaps serve as an 
attempt to compute such back-reaction effects. Within the gauge/gravity correspondence, such 
effects can be modeled in terms of the deformation of the geometry due to finite density
string cloud. This work shows that the back-reacted gravity background is explicitly computable.
There are several works, starting with  \cite{Herzog:2006gh}, where drag force on an external 
quark has been calculated by introducing D7-brane as a probe in D3 background. 
External quark on the D7-brane comes from the end point of an open string stretching
between the D7 and the horizon of the D3-brane.
The usual probe approximation here 
is justified because
the free energy of the external quark goes as ${\cal{O}}(N_c)$ whereas 
the plasma, being in adjoint representation, contributes ${\cal{O}}(N_c^2)$
to the free energy. So, in this sense, in the large colour limit, with 
$N_c \rightarrow \infty$, external quark can be treated as a probe. 
However, when large number of external quarks are introduced, the
background geometry may get modified as our work indicates. 
Before we go into our computations, here is a note of caution. We study the motion of an
external quark in an uniformly dense external quark cloud. However, in general, this motion is expected
to back-react and change the uniformity. Such effects will be neglected in this work and, therefore, 
it is within this approximation that our results
should be interpreted.

\bigskip

\noindent{\bf Gravity dual for external quark cloud}

\bigskip

We consider the $(n + 1)$ dimensional gravitational action given by
\begin{equation}
\mathcal{S} = \frac{1}{16 \pi G_{n+1}} \int dx^{n+1} {\sqrt {-g}}( R - 2 \Lambda ) + S_m,
\label{totac}
\end{equation}
where $S_m$ represents the matter part of the action. We represent the 
matter part as
\begin{equation}
S_m = -{\frac{1}{2}} \sum_i {\cal{T}}_i \int d^2\xi {\sqrt{-h}}
h^{\alpha \beta} \partial_\alpha X^\mu \partial_\beta X^\nu g_{\mu\nu},
\label{matac}
\end{equation}
where we considered $g^{\mu\nu}$ and $h^{\alpha \beta}$ are the space-time metric
and world-sheet metric respectively with $\mu, \nu$ represents 
space-time directions and $\alpha, \beta$ stands for world sheet coordinates. $S_m$ is a sum over all the string contributions
with $i$'th string having a tension ${\cal{T}}_i$. The integration in ({\ref{matac}})
is over the two dimensional string coordinates.

Varying this action with respect to the space-time metric leads to
\begin{equation}
R_{\mu\nu} - {\frac{1}{2}}R g_{\mu\nu} +\Lambda g_{\mu\nu} = 8 \pi G_{n+1} 
T_{\mu\nu},
\label{eins}
\end{equation}
with 
\begin{equation}
T^{\mu\nu} = - \sum_i {\cal{T}}_i \int d^2 \xi 
{\frac{1}{\sqrt{|g_{\mu \nu}|}}} {\sqrt{|h_{\alpha \beta}|}}
h^{\alpha \beta} \partial_\alpha X^\mu \partial_\beta X^\nu \delta_i^{n+1} 
(x - X).
\label{tmunu}
\end{equation}
In the above, the delta function represents the source divergences 
due to the presence of the strings. In the following, we will consider the
space-time metric of the form
\begin{equation}
ds^2  = g_{tt}(r) dt^2 + g_{rr}(r) dr^2 + r^2 \delta_{ab} dx^a dx^b,
\label{genmet}
\end{equation}
where $(a,b)$ run over $n-1$ space directions. We will further consider
strings with uniform tensions $T$ and use the static gauge $t = \xi^0, r = 
\xi^1$. The non vanishing components of $T^{\mu\nu}$, following from  (\ref{tmunu}), are 
\begin{equation}
T^{tt} = - \frac{a g^{tt}}{r^{n-1}}, ~~~ T^{rr} = - \frac{a 
g^{rr}}{r^{n-1}}.
\label{emten2}
\end{equation}
Here we have assumed that the strings are uniformly distributed over
$n-1$ directions such that the density is\footnote{To define this properly, we need to
think of an IR cutoff in $n-1$ directions.}
\begin{equation}
a(x) = T \sum_i  \delta_i^{(n-1)}(x - X_i),~~~{\rm with} ~a > 0.
\label{denfun}
\end{equation} 
For negative $a$, $T_{\mu\nu}$ will cease to satisfy the weak and the 
dominant energy conditions\footnote{For earlier discussions on string cloud/fluid models
see \cite{Letelier:1979ej, Stachel:1980zr, Gibbons:2000hf}.}.
We look for a solution of (\ref{eins}) in AdS space and parametrize the metric acccordingly treating $a$ as a constant\footnote{
Clearly $a$ in (\ref{denfun}) depends on $x$. However, in equations (\ref{bhole}), (\ref{t-t})
and in (\ref{comp}), $a$ is treated as constant. To do this, we have replaced $a(x)$ by an average
density as 
\begin{equation}
a = \frac{1}{V_{n-1}} \int a(x) d^{n-1} x = \frac{T}{V_{n-1}} \sum_{i=1}^{N} \int \delta_i^{(n-1)}(x - X_i) 
d^{n-1} x 
= \frac{T}{V_{n-1}} \sum_{i=1}^N 1 = \frac{TN}{V_{n-1}}.
\end{equation}
Here, $V_{n-1}$ is the volume in $n-1$ dimensional space after imposing an IR 
cut-off.
Now we consider the limit $V_{n-1}$ going to infinity along with the number of strings N, keeping $N/V_{n-1}$  constant.},
\begin{equation}
ds^2  = - V(r) dt^2 + \frac{dr^2}{V(r)} + r^2 h_{ij} dx^i dx^j.
\label{bhole}
\end{equation}
Here $h_{ij}$ is the metric on the $(n-1)$ dimensional boundary.
 As for the matter part we will focus on to the string cloud for which the nonzero 
$T_{\mu\nu}$  components are\footnote{It turns out that replacement of $\delta_{ij}$ by 
$h_{ij}$ in (\ref{genmet}) keep the components of the stress-tensor same.}
given by
\begin{equation}
T^{t}_{t} = T^{r}_{r} = - \frac{a}{r^{n-1}}, ~~{\rm with} ~~a > 0.
\label{t-t}
\end{equation}

The solution which satisfy the Einstein's equation can be easily constructed.
It is given by\footnote {This is a slight generalization of the metric in 
\cite{Herscovich:2010vr}.} 
\begin{equation}
V(r) = K + \frac{r^2}{l^2} - \frac{2 m}{r^{n-2}} - \frac{2 a}{(n-1) r^{n-3}}.
\label{comp}
\end{equation}
Here $K = 0, 1, -1$ depending on whether the $(n-1)$ dimensional boundary is 
flat, spherical or hyperbolic respectively,  having curvature $(n-1)(n-2) K$ and volume 
$V_{n-1}$. In writing down $V(r)$ we have also
parametrized cosmological constant as $\Lambda = - n(n-1)/(2 l^2)$. With equation (\ref{comp}),
the metric (\ref{bhole}) represents a black hole with singularity at $r=0$ and the
horizon is located at $V(r) =0$. The horizon has a topology of flat, spherical or hyperbolic
depending on the value of $K$. However, our interest in this work, lies in the $K =0$ case. In this case of flat horizon, 
the integration constant $m$ is related to the ADM ($M$) mass of the black hole
as follows,
\begin{equation}
M = \frac{(n-1) V_{n-1} m}{8 \pi G_{n+1}}.
\end{equation}
The horizon radius, denoted by $r_+$,
satisfies the following equation
\begin{equation}
 \frac{r_+^2}{l^2} - \frac{2 m}{r_+^{n-2}} - \frac{2 a}{(n-1) r_+^{n-3}} = 0.
\label{hor}
\end{equation}
This allow us to write $m$ in terms of horizon radius as
\begin{equation}
m = \frac{(n-1) r_+^n - 2 a l^2 r_+}{2 (n-1) l^2}.
\label{mas}
\end{equation}

The temperature of the black hole is given by
\begin{equation}
T = \frac{ {\sqrt{g^{rr}}} \partial_r {\sqrt{g_{tt}}} }{2 \pi}|_{r = r_+} 
= \frac{ n(n-1)r_+^{n+2} - 2 a l^2 r_+^3}{4 \pi (n-1) l^2 r_+^{n+1}}.
\label{flattemp}
\end{equation}
Note that the zero mass black hole has a non-zero temperature and is 
given by
\begin{equation}
T_0 = \frac{a}{2\pi} \Big( \frac{n - 1}{2 a l^2}\Big)^{\frac{n-2}{n-1}}.
\label{tempzero}
\end{equation}
The black hole temperature 
increases with the horizon size and for large $r_+$, it behaves as $T \sim r_+/l^2$.
The entropy is defined as 
\begin{equation}
S = \int T^{-1} dM,
\end{equation}
leading to the entropy density 
\footnote{Due to the nature of $a$ dependent term in (\ref{comp}), our definition of ADM mass is perhaps ambiguous. However, with this definition,
entropy of the black hole comes out to be one quarter of the horizon area, provided we assume that the equation $dS = MdT$ holds for 
this black hole. }
\begin{equation}
s = \frac{r_+^{n-1}}{4 G_{n+1}}.
\end{equation}
Note that $s$ is finite even for black hole with zero mass.
The specific heat associated with the black hole is
\begin{equation}
C = \frac{\partial M}{\partial T} = 
\frac{V_{n-1}(n-1)r_+^{n-1}(n (n -1) r_+^n - 2 a l^2 r_+)}{4 G_{n+1}  (n (n-1) r_+^n + 
2 (n-2) a l^2 r_+)}.
\label{sh}
\end{equation}
Now we have a detail look at the thermodynamic quantities just evaluated. 
First of all, if we restrict the temperature to be non-negative, the black hole
can have minimum radius
\begin{equation}
r^{{\rm min}}_+ = \Big(\frac{2 a l^2}{n(n-1)}\Big)^{\frac{1}{n-1}}.
\label{rmin}
\end{equation}
It can easily be checked that if we focus on to the region $T \ge 0$, there is
a single positive real solution of (\ref{hor}).
We also note from (\ref{rmin}) and (\ref{mas}) that the mass becomes negative at 
zero temperature
\begin{equation}
m^{{\rm min}} = -\frac{a}{n} r^{{\rm min}}_+.
\end{equation}
This is somewhat similar to the AdS-Schwarzschild with negative curvature horizon
\cite{Vanzo:1997gw}. We note that, for mass $m \ge m^{\rm min}$, the specific heat (\ref{sh})
continues to be positive and is continuous as a function of $r_+$. This suggests thermodynamical 
stability of the black hole. Finally, we write down the free energy of this black hole
\begin{equation}
{\cal F} = -\frac{(n-1) r_+^n + 2 a l^2 (n -2) r_+}{16 \pi (n-1)}.
\label{f-e}
\end{equation}

Before we go on to analyze gravitational stability of the black hole, we would like to make 
the following comment. Quite naturally, one may wonder if this black hole has a higher dimensional
origin. In particular, can this arise, in some near horizon limit,  from some brane configuration in 
ten or eleven dimensions after compactifying on spheres (with the cloud smeared over the compact manifold)?
We indeed tried to get this from some bound state configurations of D-branes and strings but
have not succeeded yet.

\bigskip

\noindent{\bf Stability of the flat black hole}

\bigskip

We now study the stability of the $K =0$ black hole geometry using the gravitational perturbation in a gauge 
invariant way \cite{Kodama1:2011ij,Kodama2:2003kj,
Kodama4:2003mj,Kodama5:2000sj,Kodama6:2001sj}. We consider perturbation on a background space time $ M^{2+p} $ 
\begin{equation}
M^{2+p} = {\mathcal{ N }}^2 \times {\mathcal{K}}^p,
\end{equation}
where the space time metric is,
\begin{eqnarray}
ds^2 = - f(r) dt^2 + \frac{dr^2}{f(r)} + r^2 \delta_{ij} dx^i dx^j, 
\nonumber \\
f(r)= \frac{r^2}{l^2} - \frac{2 m}{r^{p-1}} - \frac{2 a}{p r^{p-2}}.
\end{eqnarray}
We identify ${\mathcal{ N }}^2$ as a two dimensional space time coordinatized by t and r, whereas  ${\mathcal{K}}^p$ is a $p$ dimensional 
maximally symmetric space coordinatized by $x^{i}s$. Each perturbed tensor realized on ${\mathcal{K}}^{p}$ can be grouped into scalar, vector, 
and tensor components such that Einstein equations of motion respect the decomposition. Here we do stability analysis for tensor and 
vector perturbations. Scalar perturbation is somewhat more involved and will be reported else where.
\bigskip

\underline{Tensor perturbation}

\bigskip

In the case of the tensor perturbation, the metric tensor and energy 
momentum tensor become decomposed in scalar, vector, tensor part with respect to  ${\mathcal{K}}^{p}$ in the following manner 
\cite{Kodama2:2003kj},
\begin{eqnarray}
h_{ab} = 0, h_{ai} = 0, h_{ij}=2r^2 H_T \T_{ij}
\nonumber
\\
\delta {T}_{ab} = 0, \delta { T}^a_{i} = 0, \delta { T}^i_{j} = \tau_T \T^i_{j}, 
\end{eqnarray}
$\T_{ij}$ is the tensor harmonic function defined on ${\mathcal{K}}^{p}$. It satisfies the following properties,
\begin{eqnarray}
(\hat\triangle + k_T^2)\T_{ij} = 0
\nonumber
\\
\T^i_i = 0, \hat D_j \T^j_i = 0.
\label{tenpert1}
\end{eqnarray}
Here we note that in ${\mathcal{K}}^{p}$ space, $\hat \triangle$ and $\hat D_j$ are realized as the Laplace-Beltrami self-adjoint operator and 
the covariant derivative respectively. For $K = 0$, $ k_T^2 $ can take non-negative real continuous values \cite{Kodama4:2003mj}. 
Gauge invariant quantities like $ H_T $ and $\tau_T  $ are  function of variables belong to ${\mathcal{ N }}^2$ spacetime. \cite{Kodama5:2000sj}.
\\
Now substituting all the variations in the perturbed Einstein equation, we get the master equation of tensor perturbation \cite{Kodama2:2003kj}.
\begin{equation}
\Box H_T + \frac{p}{r}Dr \cdotp DH_T - \frac{k_{T}^2}{r^2} H_T = -{\kappa}^2 \tau_T
\label{master}
\end{equation}
We introduce a new variable $ \Phi $,
\begin{equation}
\Phi = r^{p/2} H_T,
\label{var}
\end{equation}
and substitute it into the master equation. It takes following canonical form,
\begin{equation}
\Box \Phi - \frac{V_T \Phi}{f} = 0,
\label{tenscaneqn}
\end{equation}
where $V_T$ is defined as,
\begin{equation}
V_T = \frac{f}{r^2} [{k_T}^2 +\frac{prf^{\prime}}{2}+\frac{p(p-2)f}{4}].
\label{tenspot}
\end{equation}
According to (\ref{tmunu}) the energy momentum tensor is constructed with the 
spacetime vector $X^{\mu}(t,r)$ which does not contribute to the linear
order of gauge invariant tensor perturbation. Therefore we set $\tau_T$ to be zero in (\ref{tenscaneqn}). It is clear in
plot-1 of figure (\ref{fig12}) that for higher dimensions $V_T$ is always positive beyond horizon. So the black hole geometry is stable under
tensor perturbation.
%
%
%

\pagebreak

\underline{Vector perturbation}

\bigskip

In the case of vector perturbation, metric and energy momentum tensor are decomposed in terms of vector harmonics $V_i$ as well as 
vector harmonic tensor $V_{ij}$\cite{Kodama2:2003kj}.
\begin{eqnarray}
h_{ab}=0, h_{ai}=rf_a V_i, h_{ij}=2r^2 H_T V_{ij}
\nonumber
\\
\delta {T}_{ab}=0,\delta { T}^a_{i}=r\tau^a V_i,  \delta {T}^i_{j}= \tau_T V^i_{j}
\label{vectormetric}
\end{eqnarray}
The vector harmonics are defined as 
\begin{eqnarray}
(\hat{\Delta} + {k_V}^2) V_i = 0,  \hat{D}_i V^i=0
\label{vectorharm}
\end{eqnarray}
>From vector harmonics we can construct vector type harmonic tensor,
\begin{eqnarray}
V_{ij}=-\frac{1}{2k_V}(\hat{D}_iV_j+\hat{D}_jV_i),
\nonumber
\\
( \hat{\Delta} + k_V^2)V_{ij} = 0,
 V^i_i=0, \hat{D}_j V^j_i = \frac{k_V}{2}V_i.
\label{vectorharmtens}
\end{eqnarray}
The gauge invariant parameters for K $ = $ 0 are given by
\begin{equation}
F_a = f_a + r D_a (\frac{H_T}{k_V}), \tau_T , \tau^a.
\end{equation}
Upon considering the perturbations of the Einstein equation and the conservation law of energy momentum tensor, master equation arising
from the gravitational perturbation with the source term takes the following form \cite{Kodama4:2003mj},
\begin{equation}
r^p D_a(\frac{1}{r^p}D^a\Omega) - \frac{{k_V}^2}{r^2}\Omega = -\frac{2 \kappa^2}{{k_V}^2}r^{p}\epsilon^{ab}D_a(r\tau_b)
\end{equation}
where,
\begin{equation}
r^{p-1}F^a = \epsilon^{ab}D_b\Omega + \frac{2 \kappa^2}{{k_V}^2}r^{p+1}\tau^a.
\end{equation}
Now introducing the change of variable
\begin{equation}
\Phi = r^{-p/2} \Omega
\end{equation}
we recast the master equation in a canonical form, where the effective potential for vector perturbation comes out as
\cite{Kodama6:2001sj}, 
\begin{equation}
V_V = \frac{f}{r^2}[{k_V}^2 + \frac{p(p+2)}{4}f - \frac{pr}{2}f^{\prime}]
\end{equation}
%
%

\begin{figure}[ht]
\centering
\mbox{\subfigure[Plot-1]{\includegraphics[width=5 cm]{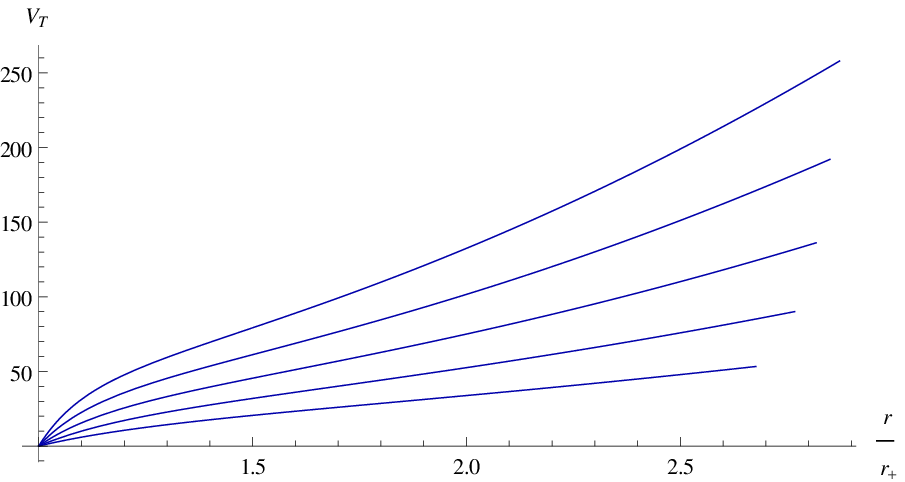}}
\quad
\subfigure[Plot-2]{\includegraphics[width=5 cm]{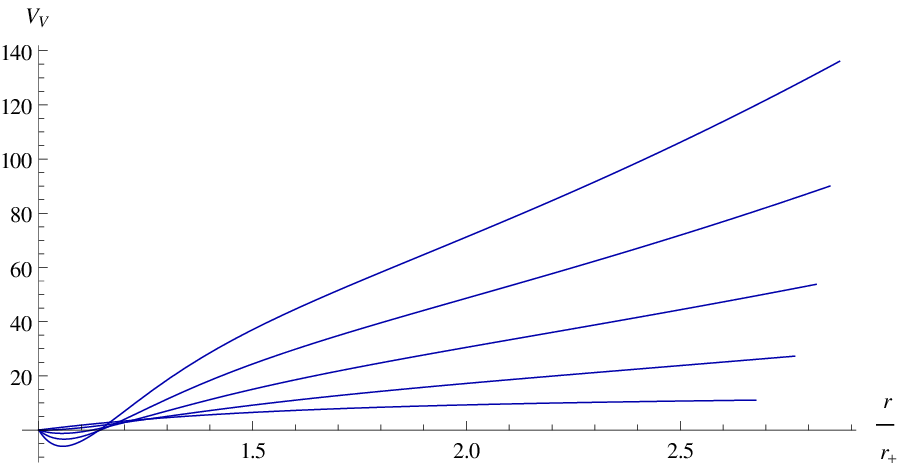} }}
\caption{Plot 1 shows, for various dimensions, the effective potential $V_T$ in tensor perturbation is positive beyond horizon radius 
. Plot 2 shows the effective potential  $V_V$ in vector perturbation is not always non-negative for $p>3$. In both cases horizontal axis is 
normalized with respect to  black hole horizon radius $r_+$.}
\label{fig12}
\end{figure}
The plot-2 in figure (\ref{fig12}) implicates that  beyond horizon, $V_V$ is not always non-negative for $p>3$. We follow $ S-deformation $
 method \cite{Kodama2:2003kj} to construct modified effective potential 
\begin{equation}
\tilde{V}_{V} = V_{V} + f\frac{dS}{dr} - S^2,
\end{equation}
where $S$ is an arbitrary function of $r$. If we choose  $ S = \frac{pf}{2r} $, we get the modified effective potential, 
\begin{equation}
{\tilde{V}}_V  = \frac{f{k_V}^2}{r^2} > 0
\end{equation}
Once again $ {k_V}^2 $ is the eigenvalue of a positive operator. So the above form of ${\tilde{V}}_V  $ furnishes the sufficient condition
for the stability of the black hole. \\
Having constructed this black hole geometry we compute the drag force on an external quark moving in external quark cloud

\bigskip

\noindent{\bf Dissipative force on an external quark moving in the heavy quark cloud}

\bigskip

We now like to calculate the dissipative force experienced by the external heavy quark 
moving in the cloud of heavy quarks. Our aim is to study the force as a function of
the cloud density. Calculational procedure to evaluate drag force on an external quark 
is by now standard. This can be found, for example, in \cite{Herzog:2006gh, Liu:2006ug, 
Gubser:2006bz}. We will follow the notations of \cite{Gubser:2006bz}. The drag force on a 
very massive quark with fundamental $SU(N)$ charge at finite temperature 
is calculated holographically by studying the motion of a string whose end point is
on the boundary. This end point represents
the massive quark whose mass is proportional to the length of the string. We will consider here
the gauge theory on $R^3$ coordinatized by $x^1, x^2, x^3$. This means, for the purpose of
this computation, we only consider $K=0, n = 4$ case of the black holes discussed previously.

Let us consider the motion of a string only in one direction, say $x^1$. In static gauge, $t = \xi^0, 
r = 
\xi^1$, the embedding of the world-sheet is given by the function $x^1(t,r)$. The induced action 
of the string in our case follows from a straightforward computation
\begin{equation}
S = -{\frac{1}{2 \pi \alpha^{\prime}}} \int dt dr {\sqrt{   1 + \frac{3 r^4 - 2 a l^2 r - 6 m l^2}{3 l^4} (\partial_r x^1)^2 
- \frac{3 r^4}{3 r^4 - 2 a l^2 r - 6 m l^2} (\partial_t x^1)^2 }},
\label{ss}
\end{equation}
where we have scaled $x^1$ by $l$.

The ansatz that describes the behaviour of the string with attached quark moving 
with constant speed $v$ along $x^1$ is given by \cite{Gubser:2006bz}
\begin{equation}
x^1 (r, t) = v t + \xi(r),
\end{equation}
for which (\ref{ss}) simplifies to
\begin{equation}
S = -{\frac{1}{2 \pi \alpha^{\prime}}} \int dt dr {\sqrt{   1 + \frac{3 r^4 - 2 a l^2 r - 6 m l^2}{3 l^4}(\partial_r \xi)^2
- \frac{3 r^4}{3 r^4 - 2 a l^2 r - 6 m l^2} v^2  }}.
\end{equation}
The momentum conjugate to $\xi(r)$ is 
\begin{equation}
\pi_\xi = -{\frac{1}{2 \pi \alpha^{\prime}}} \frac{(3 r^4 - 2 al^2 r - 6 m l^2) \partial_r \xi}{ {3 l^4
\sqrt{   1 + \frac{3 r^4 - 2 a l^2 r - 6 m l^2}{3 l^4} (\partial_r \xi)^2- \frac{3 r^4}{3 r^4 - 2 a l^2 r - 6 m l^2}  v^2  }}}.
\end{equation}
Equation of motion can be obtained by inverting this equation for $\partial_r \xi$.
However, as in \cite{Gubser:2006bz}, to get a real $\xi$, the constant of motion $\pi_\xi$ 
has to be set to 
\begin{equation}
\pi_\xi = -{\frac{1}{2 \pi \alpha^{\prime}}} \frac{v r_v^2}{l^2}
\label{conmom}
\end{equation}
where $r_v$ is the real positive solution of the equation
\begin{equation}
3(1 - v^2) r_v^4 - 2 a l^2 r_v - 6 m l^2 = 0.
\label{const}
\end{equation}
Though this equation can be solved explicitly, the solutions are not very illuminating.
However, it is easy to check that there is only one real positive solution. Substituting
this solution of $r_v$ in (\ref{conmom}), we get $\pi_\xi$. The dissipative force is then given by
\cite{Gubser:2006bz}
\begin{equation}
F = -{\frac{1}{2 \pi \alpha^{\prime}}} \frac{v r_v^2}{l^2},
\label{dissi}
\end{equation}
with $r_v$ given by the positive real solution of (\ref{const}). \*
Now we wish to rewrite the expression of the dissipative force in terms of gauge theory parameters. Along this line, we solve (\ref{flattemp}) for $r_+$,
\begin{equation}
r_{+} = \frac{l^2}{6} A(T,b),
\label{solvetemp}
\end{equation}

where $b$ is the scaled quark cloud density, $b = a/l^4$ and $A$ is given by,
\begin{eqnarray}
A(T,b) & = & [{\{2 (9 b+4 \pi ^3 T^3)+6\sqrt{b(9b + 8 \pi^3 T^3)}\}}^{1/3} + \nonumber \\
& & 2\pi T \{ 1 + \frac{2^{2/3}\pi T } {\{(9b + 4 \pi^3 T^3) + 3 \sqrt{b(9 b+8 \pi ^3 T^3)}\}^{1/3} } \}]
\label{relation2}
\end{eqnarray}\\*
Substituting (\ref{solvetemp}) and the following useful relation
\begin{equation}
\frac{l^4}{{\alpha^{\prime}}^2} = {g_{YM}}^2N, 
\label{relation1}
\end{equation}
in the expression of the dissipative force (\ref{dissi}), we get the modified form
\begin{equation}
 F = -\frac{A^2}{72\pi}\sqrt{{g_{YM}}^2 N} v \frac{{r_v}^2}{{r_{+}}^2}.
\label{gauge}
\end{equation}
Here $g_{YM}$ is the Yang-Mills(YM) gauge coupling and $N$ is the order of the gauge group $SU(N)$.
We are able to solve the ratio $ {r_v}^2/{r_{+}}^2$ in a closed form by substituting (\ref{hor}) 
into (\ref{const}). The relevant equation takes the following form,
\begin{equation}
(1-v^2)(\frac{{r_v}^4}{{r_{+}}^4}) - \frac{144b}{{(A(T,b))}^3}((\frac{r_v}{r_{+}})-1)-1 = 0,
\label{finalform}
\end{equation}
It turns out that the real positive solution of (\ref{finalform}) is expressible in terms of $A(T,b)$ and $b$ itself. Denoting the solution as $f(A,b)$ and plugging it back into (\ref{gauge}) we achieve the form
of dissipative force expressible in terms of gauge theory parameters
\begin{equation}
 F = -\frac{A^2}{72\pi}\sqrt{{g_{YM}}^2 N} v f(A,b)^2.
\label{gauge1}
\end{equation}
We note here that $f(A,b)$ is an explicitly computable function.

\begin{figure}
\centering
\mbox{\subfigure[Plot 1]{\includegraphics[width=6.5 cm]{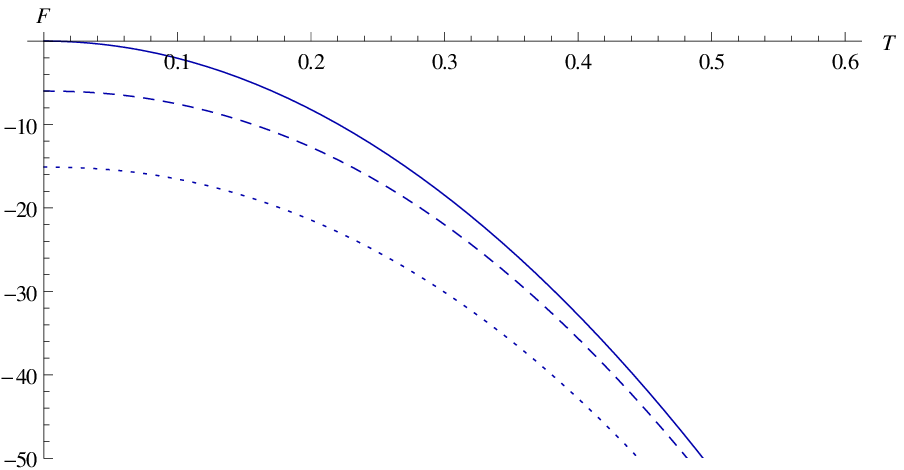}}
\quad
\subfigure[Plot 2]{\includegraphics[width=6.5 cm]{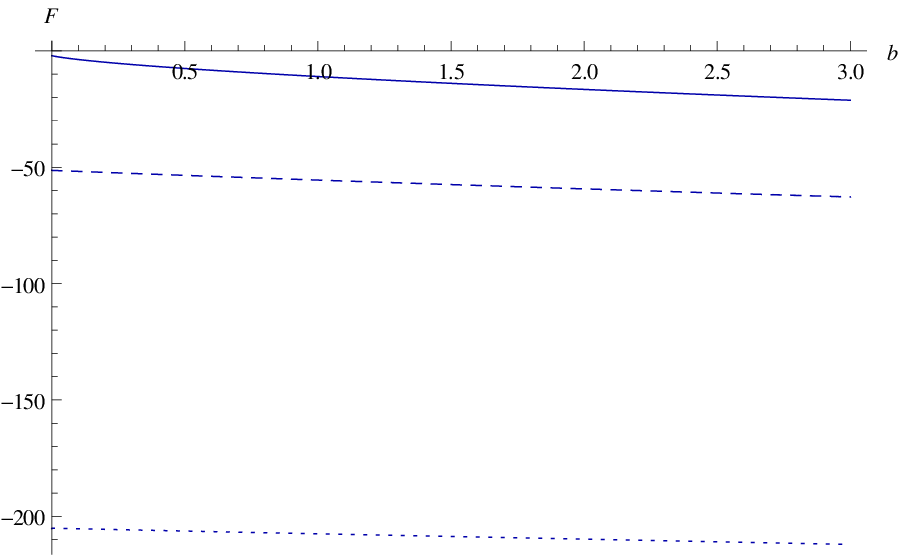} }}
\caption{Plot 1 shows the variation of $F$ as a function of $T$ for the
 values of quark density $b = $ 0 (solid), 0.5 (dashed), 2 (dotted) respectively. Plot 2 shows the variation of 
$F$ as a function of $b$ for the
values of $T= $ .1 (solid), .5 (dashed), 1 (dotted) respectively. We see in both cases the larger the quark cloud density as well 
as temperature, the more is the dissipative force.}
\label{fig12}
\end{figure}

We would now study (\ref{gauge1}) for different values of heavy quark density and for fixed $T$.
As for an example, it is interesting to check that if the temperature is fixed at the value $T_0$ as mentioned in (\ref{tempzero})
the dissipative force behaves as $F \sim - b^{2/3}$, where $b$ is now the
density of the quark cloud. Also for $T = 0$, $A(T,b)$ in (\ref{relation2}) simplifies significantly resulting
the dissipative force to vary as $F \sim - b^{2/3}$. For generic temperature and small $b$, it is possible
to have a power series solution of (\ref{gauge1}) in $b$. However, for appreciable
density, we find it more suitable to analyze $F$ in terms of plots.
In figure (\ref{fig12}) plot 1 shows the behaviour of the drag force as a function of
$T$ for different $b$. For fixed $T$, we clearly see that
the force becomes stronger with the quark density\footnote{Note that the free energy (\ref{f-e}) is 
perfectly well behaved at $T =0$. Infact, it is ${\cal F} = -\frac{3 a r_+}{32 \pi}$. Substituting $r_+$,
we find ${\cal F} \sim - b^{\frac{4}{3}}$. Furthermore computation of the drag force leads to
$F \sim - b^{\frac{2}{3}}$ in this limit.}

\bigskip

\noindent{\bf Conclusion}

\bigskip

In this paper, using the AdS/CFT duality, we have computed the dissipative force experienced by an external heavy quark with fundamental 
$ SU(N) $ charge moving in the heavy quark cloud at finite temperature. In the dual theory, we have considered motion of a string 
(with one of it's end point at boundary) in a $n+1$ dimensional  background with flat boundary pervaded with string cloud as the matter source. 

The geometry implies existence of a black hole parametrized by it's mass $ m $ and the string cloud density $ a $. The black hole turns 
out to be thermodynamically stable and it resembles AdS-Schwarzschild black hole with negative curvature horizon. We have been able to
check the gravitational stability of the geometry for tensor and vector perturbations.

In the above scenario, we have computed the drag force exerted on the external quark. The most general form of the dissipative 
force turns out to be  a complicated function of temperature of the boundary theory $ T $ and the re-scaled quark cloud density $ b$. 
However for the temperature corresponding to massless black hole in the dual gravity theory, it behaves like  $F\sim -b^{\frac{2}{3}}$. 
We have plotted it with respect to both $ T$ and $ b$ separately while keeping one of them constant at a time. 
Both plots exhibit an enhancement in the drag force in the presence of evenly distributed quark cloud.

There are few issues that we hope to look into. Firstly, what is the higher dimensional brane geometry whose 
near horizon geometry contains the black hole that
 we are considering? Secondly, is the gravitational background stable under scalar perturbation?

\bigskip

\noindent{\bf Acknowledgments}

\bigskip

I would like to thank Sudipta Mukherji for suggesting me this problem  and for useful discussions. 
I am thankful to Shibaji Roy for his  valuable comments on an initial version of this paper. I would also like to 
thank Yogesh Srivastava, Ajit M. Srivastava, Sachin Jain and Sayan Chakrabarty for 
several useful comments. I am thankful to all the other members of string group at IOP for encouragement.
Finally, I thank the referee for very useful comments which led to substantial improvement of the
manuscript.

\end{document}